\documentclass[12pt]{article}
 \usepackage{amsmath}
  \usepackage[dvips]{graphicx}
\newcommand{\nc}{\newcommand}
\nc{\non}{\nonumber}
\nc{\pt}{p_{{}_T}}
\nc{\lmc}{\Lambda_c^+}
\nc{\plmc}{\vec{\Lambda}_c^+}
\nc{\all}{A_{LL}}
\nc{\dg}{\Delta G(x,Q^2)}
\nc{\stil}{\tilde{s}}
\nc{\ttil}{\tilde{t}}
\nc{\util}{\tilde{u}}
\nc{\shs}{\hat{s}}
\nc{\ths}{\hat{t}_1}
\nc{\uhs}{\hat{u}_1}
\nc{\cosec}{\rm cosec}
\nc{\mpr}{m_p}
\nc{\mc}{m_c}
\nc{\mlc}{m_{\lmc}}
\nc{\pa}{p_{{}_a}}
\nc{\pb}{p_{{}_b}}
\nc{\pA}{p_{{}_A}}
\nc{\pB}{p_{{}_B}}
\nc{\plc}{p_{\lmc}}
\nc{\pc}{p_{{}_c}}
\newcommand{\bm}[1]{\mbox{\boldmath $#1$}}
\begin{document}
\pagestyle{empty} \setlength{\footskip}{2.0cm}
\setlength{\oddsidemargin}{0.5cm} \setlength{\evensidemargin}{0.5cm}
\renewcommand{\thepage}{-- \arabic{page} --}
   \def\thebibliography#1{\centerline{\large \bf REFERENCES}
     \list{[\arabic{enumi}]}{\settowidth\labelwidth{[#1]}\leftmargin
     \labelwidth\advance\leftmargin\labelsep\usecounter{enumi}}
     \def\newblock{\hskip .11em plus .33em minus -.07em}\sloppy
     \clubpenalty4000\widowpenalty4000\sfcode`\.=1000\relax}\let
     \endthebibliography=\endlist
   \def\sec#1{\addtocounter{section}{1}\section*{\hspace*{-0.72cm}
     \normalsize\bf\arabic{section}.$\;$#1}\vspace*{-0.3cm}}
\begin{flushright}
\vspace*{-2cm}
KOBE-FHD-02-06\\
YNU-HEPTH-02-105\\
hep-ph/0212366
\end{flushright}
\renewcommand{\thefootnote}{$\dag$}
\begin{center}
{\bf \huge
       Charmed Hadron Production in \\
 Polarized \bm{pp} Reactions as a Probe\vspace*{0.2cm}\\ 
        of Polarized Gluons in the Proton}{\Large \footnote{
  Talk presented by K. Ohkuma at
  the XVI  Particle and Nuclei International conference, \\
  \phantom{aaaa}Osaka, Sep. 30 - Oct. 4, 2002}}
\vspace*{1cm}\\

{\sc \Large Toshiyuki MORII}\vspace*{0.2cm}\\
Division of Sciences for Natural Environment,\\
Faculty of Human Development,\\
Kobe University, Nada, Kobe 657-8501, JAPAN\\
Electronic address{\tt :morii@kobe-u.ac.jp}\vspace{1.0cm}\\

{\sc \Large Kazumasa OHKUMA}\vspace*{0.2cm}\\
Department of Physics,\\
Faculty of Engineering,\\
Yokohama National University,\\
 Hodogaya, Yokohama
240-8501, JAPAN\\
Electronic address{\tt :ohkuma@phys.ynu.ac.jp}
\end{center}
\vspace*{0.5cm}

\centerline{\bf \large ABSTRACT}
\vspace*{0.2cm}
\baselineskip=20pt plus 0.1pt minus 0.1pt

To probe the behavior of polarized gluons in the proton, 
we propose the charmed hadron, such as $\lmc$ and $D^{*}$, 
production in the forthcoming RHIC experiments.
We found that the spin correlation between the target proton
and the produced $\lmc$ baryon might be a good signal for 
testing models of the polarized gluon distribution in the proton.

\newpage

\renewcommand{\thefootnote}{$\sharp$\arabic{footnote}}
\renewcommand{\thesection}{\Roman{section}.$\!\!$}
\pagestyle{plain} \setcounter{footnote}{0}
\pagestyle{plain} \setcounter{page}{1}
\baselineskip=18.0pt plus 0.2pt minus 0.1pt
\section{INTRODUCTION}
The advent of so-called ``the proton spin puzzle'' which has emerged
from the measurement of the polarized structure function of proton
$g^p_1(x)$ by the EMC collaboration~\cite{EMC}, has stimulated 
a great theoretical and experimental activities in nuclear
and particle physics~\cite{rpsp}.
Though a great deal of efforts have been made for solving 
this puzzle so far, many problems still remain to be solved.   
As is well known,
the spin of proton is carried by quarks, gluons and their 
orbital angular momenta. 
From the next--to--leading order QCD analysis for many and precise data 
on the polarized structure function $g_1(x)$ of nucleons, 
now we have a rather good knowledge on the polarized
quark distribution in the proton.    
However, the polarized gluon distribution $\Delta G(x)$ 
in the nucleon is still 
very uncertain.  To know how the gluon polarizes in the nucleon
is very important to solve the proton spin puzzle.
So far, the behavior of gluons in the proton have been studied
in many cases for deep inelastic polarized $e-p$ scattering. 
However, the Relativistic Heavy Ion Collider(RHIC) could open another 
chance to probe internal structure of proton via 
$\vec{p}\vec{p}$ collisions.
In the RHIC experiments~\cite{RHIC}, several interesting processes, such as
high $\pt$ prompt photon production,  
jet production,
heavy flavor production, etc. are proposed to probe the polarized gluon
in the proton.
Here we also propose another processes, i.e. the polarized
charmed hadron production, $p\vec{p}\to \vec{\Lambda}_c^+X$ and 
$p\vec{p}\to \vec{D}^*X$, in the polarized proton--unpolarized proton
collision to extract information about 
$\Delta G(x)$.\footnote{In this report we focus only on the $\Lambda_c^+$ 
production, though we have calculated for $D^*$ production, too, 
because the main point of the result remain unchanged.}
In these processes, $\lmc$ is mainly produced via fragmentation 
of a charm quark which is  originated dominantly 
from gluon-gluon fusion subprocess.\footnote{This is because 
charm quarks are tiny contents in the proton.}
Thus, its cross section is expected to be sensitive to the gluon
distribution in the target proton.
Moreover, since $\lmc$ is composed of a heavy charm quark and
antisymmetrically combined light up and down quarks,
the spin of $\lmc$ is basically carried by 
a charm quark which is produced via gluon-gluon fusion subprocess.
Therefore, observation of the spin of the produced $\lmc$ gives us
information about $\Delta G(x)$ in the nucleon.

\section{SPIN CORRELATION ASYMMETRY AND ITS STATISTICAL SENSITIVITY}
As a useful observable to extract $\Delta G(x)$ 
in the proton, 
we introduce a spin correlation asymmetry between the target 
polarized--proton and
produced $\Lambda_c^+$ baryon;
\begin{eqnarray}
&&A_{LL}=\frac{d \sigma_{++} - d\sigma_{+-} + d \sigma_{--}- d\sigma_{-+}}
{d \sigma_{++} + d \sigma_{+-}+d \sigma_{--} + d \sigma_{-+}}
\equiv
\frac{{d \Delta \sigma}/{d X}}
{{{d\sigma}/{d X}}},~~(X=\pt~{\rm or}~\eta), \label{eq:all}
\label{all}
\end{eqnarray}
where $d \sigma_{+ -}$, for example, denotes the spin-dependent
differential cross section with the positive helicity of the target proton
and the negative helicity of the produced $\Lambda_c^+$ baryon.
$\pt$ and $\eta$, which are represented as $X$ in Eq.(\ref{all}),
are transverse momentum and pseudo-rapidity of
produced $\lmc$, respectively. 
The spin-independent(-dependent) differential cross section 
$d (\Delta) \sigma /d X$ can be calculated by the quark-parton model 
(see ref.~\cite{osm} for details).


Statistical sensitivities of $\all$ for the $\pt$  and  
$\eta$ distribution are estimated by using the following formula;
\begin{equation}
\delta \all \simeq \frac{1}{P}\frac{1}{
\sqrt{b_{\Lambda_c^+}~\epsilon~L~T~\sigma}}.
\end{equation}
To numerically estimate the value of $\delta \all$, here 
we use following parameters:
operating time; $T=$100-day, the beam polarization; $P=$70\%,
a luminosity; $L=8\times 10^{31}~(2\times 10^{32}$)~cm$^{-2}$ sec$^{-1}$ for
$\sqrt{s}=200~(500)~$ GeV,
the trigger efficiency; $\epsilon =10\%$ for detecting produced
$\Lambda_c^+$ events and a branching ratio; $b_{\Lambda_c^+}\equiv 
{\rm Br}(\lmc \to p K^- \pi^+)\simeq5\%$~\cite{pdg}.
The branching ratio of this purely charged decay mode is 
needed to measure the polarization of
produced $\lmc$. 
$\sigma$ denotes the unpolarized cross section integrated over suitable
$\pt$ or $\eta$ region.
\section{NUMERICAL ANALYSIS}
In the numerical calculation of $A_{LL}$, 
we limited the integration region of $\eta$ and 
$\pt$ of produced $\lmc$  as
$-1.3 \leq \eta \leq 1.3$ and
3 GeV $\leq \pt \leq$ 15(40) GeV, respectively, 
for  $\sqrt{s}=200(500)$ GeV.  The range 
of $\eta$ and the lower limit of $\pt$ were selected 
in order to get rid of the contribution from
the diffractive $\lmc$ production. 
As for the upper limit of $\pt$, we took it as described above, 
for simplicity, though the kinematical maximum of $\pt$
of produced $\Lambda_c^+$ is 
slightly larger than 15 GeV and 40 Gev for $\sqrt{s}=$200 GeV 
and 500 GeV, respectively.
In addition, we took the AAC\cite{aac} and  GRSV01~\cite{grsv}
parameterization models for the polarized gluon distribution 
function and the GRV98~\cite{grv} model for the
unpolarized one.
Though both of AAC and GRSV01 models excellently reproduce the experimental 
data on the polarized structure function of nucleons $g_1(x)$,
$\Delta G(x)$ for those models are quite different. 
Therefore, those models should be tested in other processes. 
Since our process is semi-inclusive, the fragmentation function of a 
charm quark to $\Lambda_c^+$ is necessary to carry out numerical calculations.
For the unpolarized fragmentation function, 
we used Peterson fragmentation function, 
$D_{c \to \Lambda_c^+}(z)$~\cite{pdg,peter}.
However, since we have no data, at present, 
about polarized fragmentation
functions $\Delta D_{\vec{c}\to \vec{\Lambda}_c^+}(x)$ 
for the polarized $\Lambda_c^+$ production, we took the
following ansatz for it
\begin{equation}
\Delta D_{\vec{c}\to \vec{\Lambda}_c^+}(z)= C_{c\to \Lambda_c^+} 
D_{c \to \Lambda_c^+}(z),
\end{equation}
where $C_{c \to \Lambda_c^+}$ is a 
scale-independent spin transfer coefficient.
In this analysis, we studied two cases: (A) 
$C_{c\to \Lambda_c^+}=1$ (non-relativistic quark model) and 
(B) $C_{c\to \Lambda_c^+}=z$ (Jet fragmentation model~\cite{jet}).
As we discussed before,
if the spin of $\lmc$ is equal to 
the spin of the charm quark produced in the 
subprocess, the model (A) might be a reasonable scenario.

\section{RESULTS AND DISCUSSION}
\begin{figure}[htb]
  \includegraphics[scale=0.30]{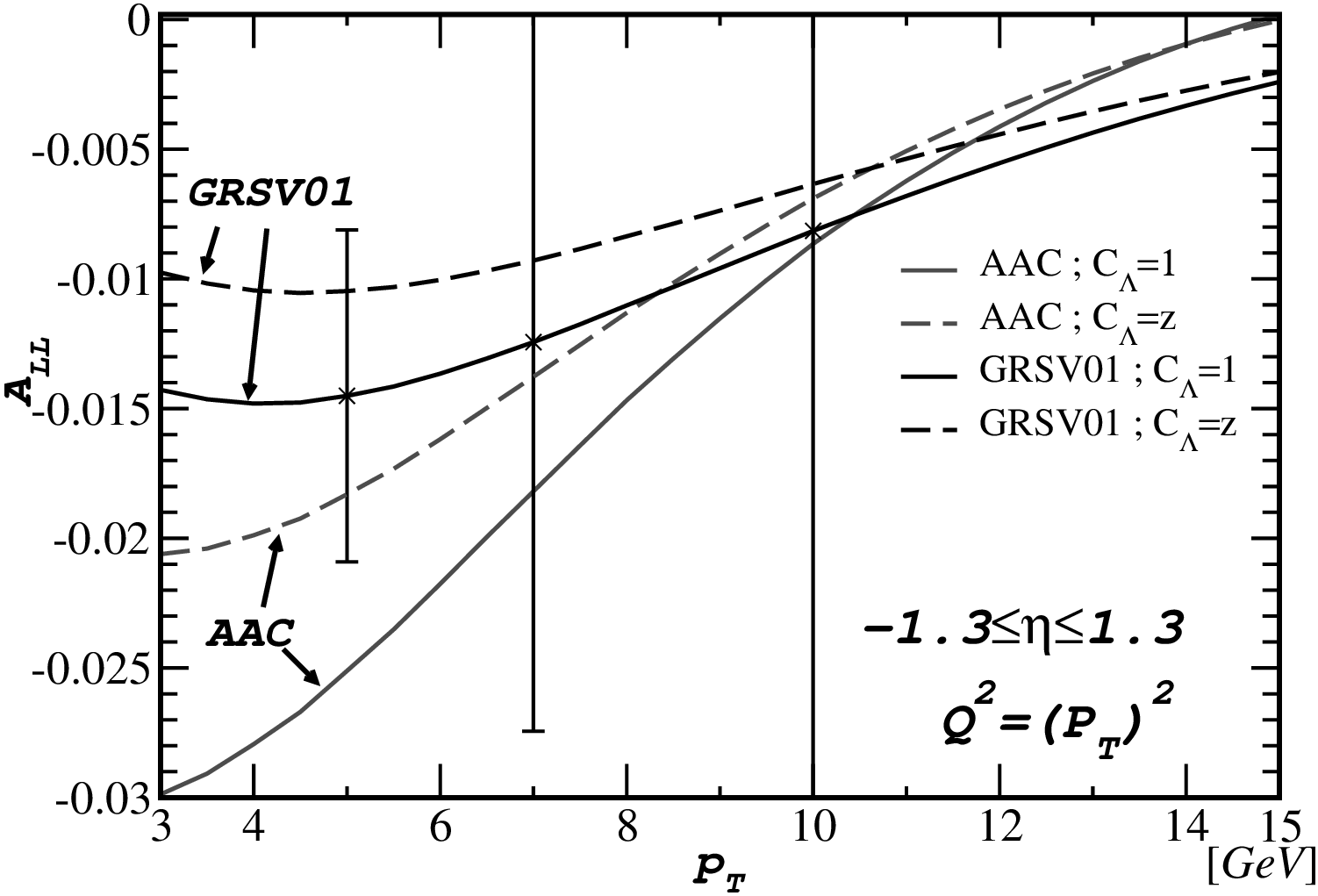}~
  \includegraphics[scale=0.30]{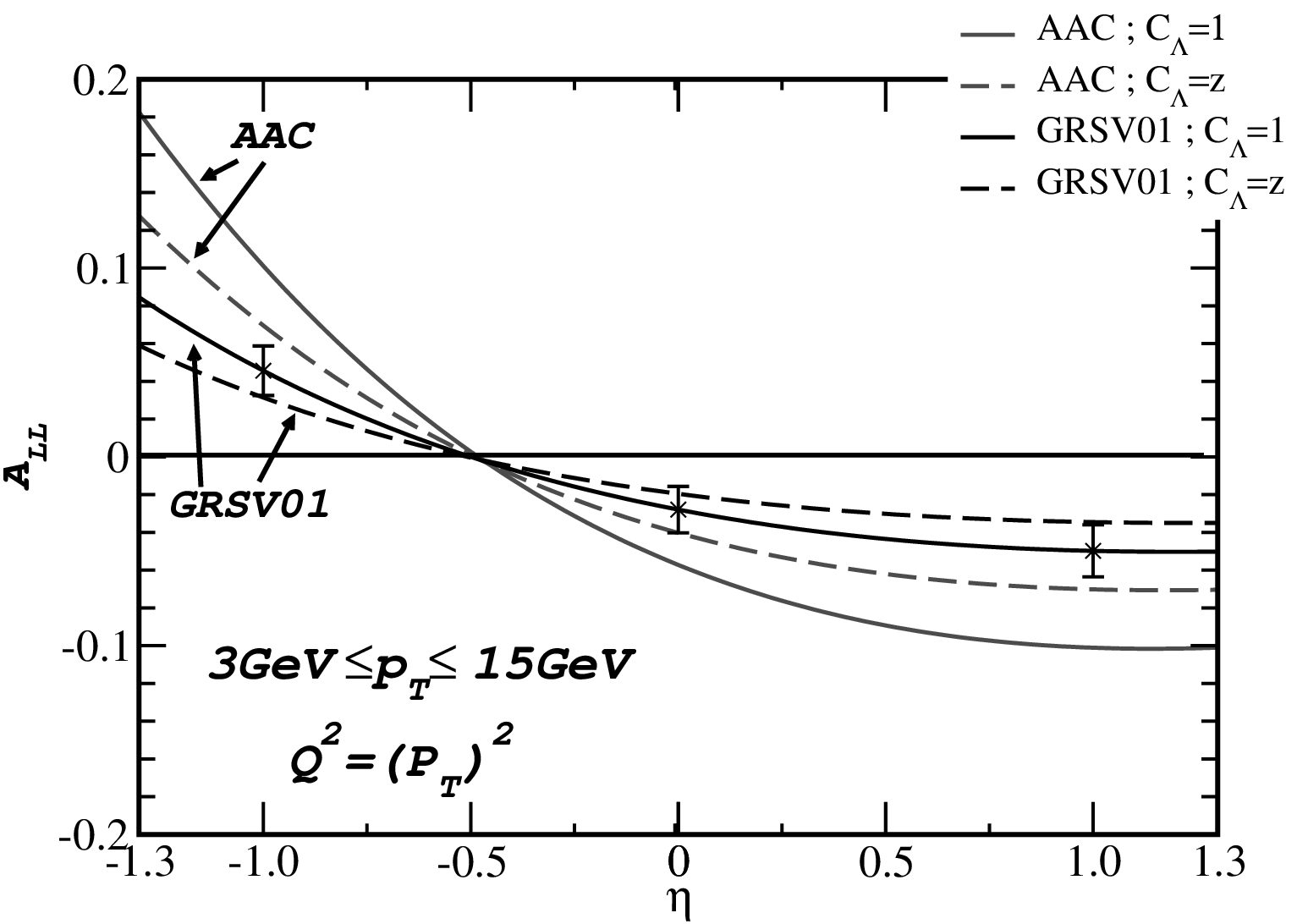}
\caption{Spin correlation asymmetry of $\pt$(left panel) and $\eta$
 (right panel) distribution at $\sqrt{s}=200$ GeV}
\label{200}
\end{figure}

\begin{figure}
  \includegraphics[scale=0.30]{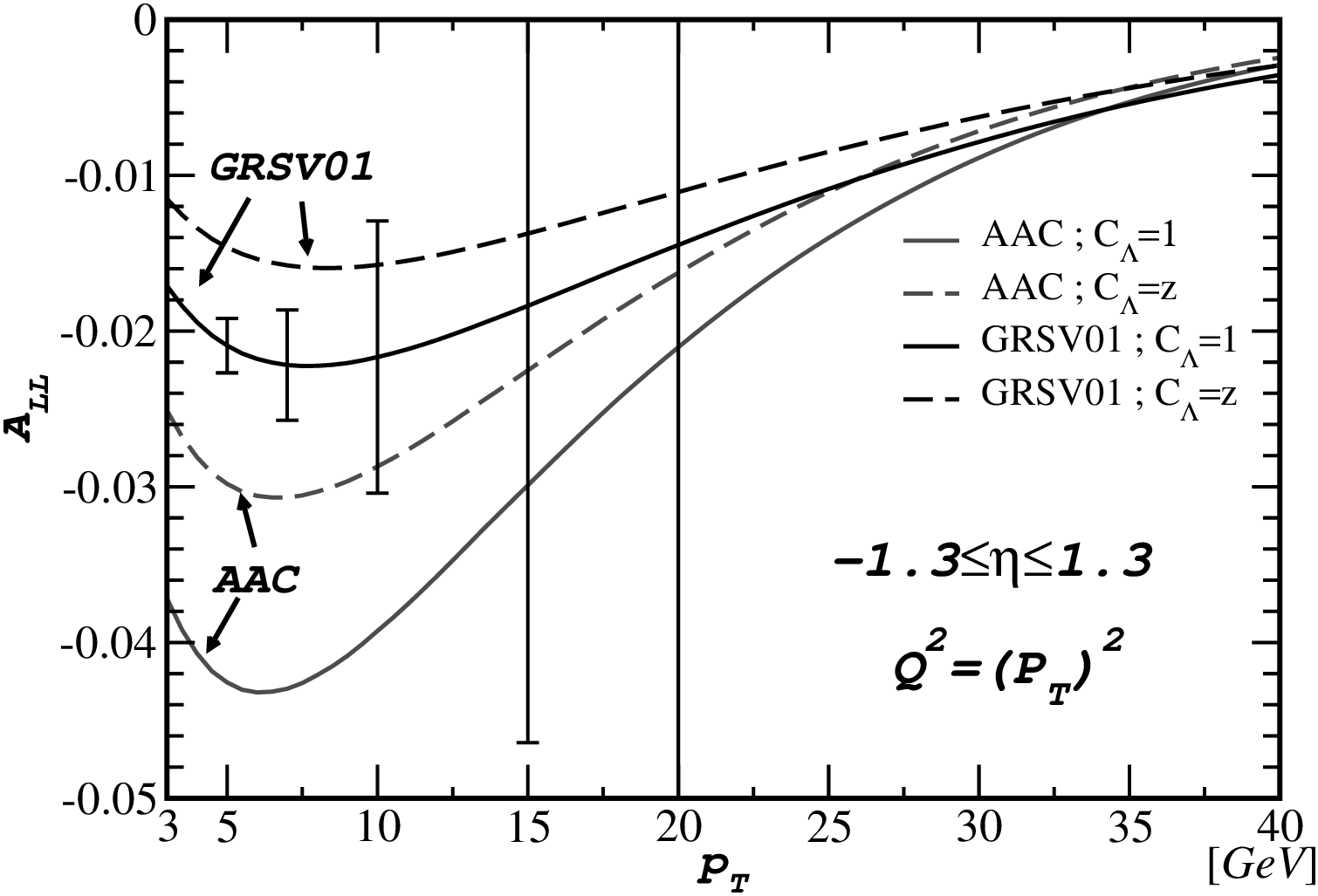}~
  \includegraphics[scale=0.30]{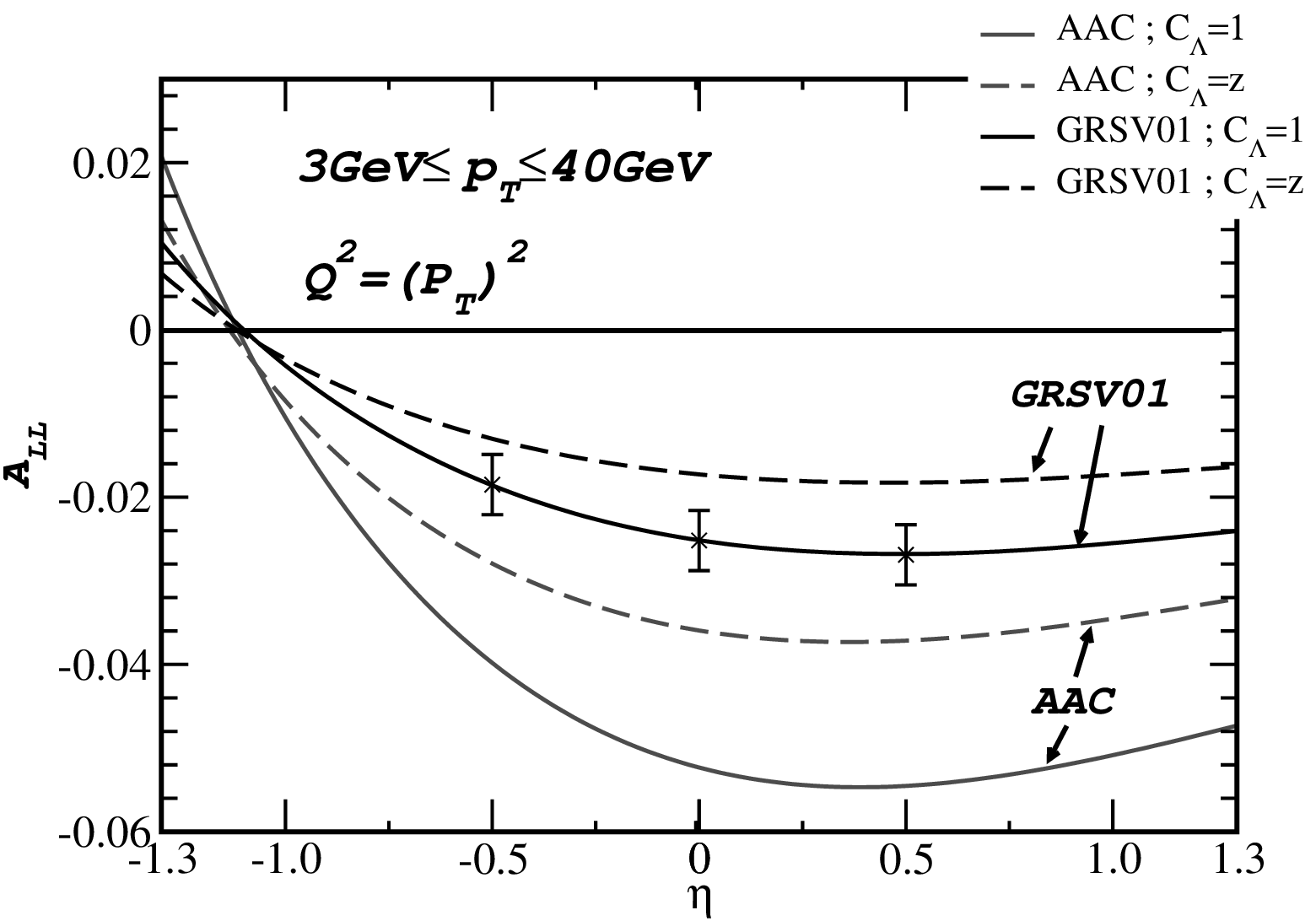}
\caption{The same as in Fig.~\ref{200}, but for $\sqrt{s}=500$ GeV}
\label{500}
\end{figure}

Numerical results of $\all$ are shown in Fig.~\ref{200} 
and Fig.~\ref{500}. 
In these figures, we attached $\delta \all$ 
to the solid line of $\all$  calculated for the 
case of the GRSV01 parametrization model of 
polarized gluon and the non-relativistic 
fragmentation model.\footnote{Note that as shown 
from Eq.(2), $\delta A_{LL}$ 
does not depend on both of the model of polarized gluons and the model 
of fragmentation functions.} 
Comparing with  those figures, we can see that
the $\eta$ distributions of $\all$ are more effective 
than the $\pt$ distributions at $\sqrt{s}=$200 GeV and 500 GeV
to distinguish various models.
%
%
As shown in the right panel of Fig.~\ref{500} given at $\sqrt{s}=$500 GeV,
we could distinguish the parametrization models of 
polarized gluon as well as the models of the spin-dependent 
fragmentation function, though the magnitude of $\all$ is rather small.
At $\sqrt{s}=$200 GeV, the magnitude of $\all$ for $\eta$ distribution 
becomes larger, though statistical sensitivities
are not so small.
If the integrated luminosity at  $\sqrt{s}=$200 GeV becomes 
large and the detection efficiency $\epsilon$ is improved,
this observable could be promising to distinguish
not only the models of  $\Delta G(x)$ but also
the models of $\Delta D(z)$.
On the other hand, we cannot say anything from 
the $\pt$ distribution of $\all$ at high $\pt$ region, 
since $\delta \all$ becomes rapidly large with increasing $\pt$.
However, if we confine the kinematical region in 
rather small $\pt$ range such as 
$\pt=3\sim$5(10) GeV at $\sqrt{s}=200(500)$ GeV,  
it might be still effective.

Though this analysis is confined to leading order,
the results are interesting and we hope our prediction will be
tested in the forthcoming RHIC experiment.\\
%
\end{document}